\documentclass[11pt]{amsart}

\usepackage{amsmath,amsthm,amsfonts,amssymb}
\usepackage{graphicx}
\usepackage{hyperref}
\usepackage[usenames]{color}

\newtheorem{theorem}{Theorem}[section]

\theoremstyle{definition}

\newtheorem{remark}[theorem]{Remark}

\newcounter{comcount}

\numberwithin{equation}{section}

\title{
Cryptanalysis of a new version of the MOR scheme }

\author{Vitaly Roman'kov}

\address{Dostoevsky Omsk State University}
\email{romankov48@mail.ru}

\date{}

\begin{document}

\maketitle

\footnote{Supported by RFBR, project 18-41-550001.}

\begin{abstract}
We show that an attack based on the linear decomposition method introduced by the  author can be efficiently applied to the new version of the MOR scheme proposed in  \cite{BMSS}. We draw attention to some inaccuracies in the description of this version. We show how the action of an exponent of a given automorphism (for example, the action of its inverse) can be calculated,  and we also show how the  unknown exponent of automorphism can be calculated if we go over to the corresponding linear transformation. This method can be applied to different matrix groups over an arbitrary constructive field. It does not depend on the specific properties of the underlined matrix group.   The considered problem is reduced in probabilistic polynomial time to the similar problem in small extensions of the underlined field. 
\end{abstract}

\section{Introduction}

 In \cite{BMSS}, S. Bhunia, A. Mahalanobis, P. Shinde and A. Singh  study the ElGamal-type version of the MOR cryptosystem with symplectic and orthogonal groups over finite fields $\mathbb{F}_q$ of odd characteristics. The MOR cryptosystem over SL($d, \mathbb{F}_q$) was previously investigated by the second of these authors. In that case, the hardness of the MOR cryptosystem was found to be equivalent to the discrete logarithm problem in $F_{q^d}$. It is shown in \cite{BMSS} that the MOR cryptosystem over Sp($d, q$) has the security of the discrete logarithm problem in $\mathbb{F}_{q^d}.$ The MOR cryptosystem  also studied in \cite{Paeng} and in \cite{Mah}, and was cryptanalyzed in \cite{Monico}. 
 
Recall, that the El-Gamal system can be described as follows: Let $G$ be a public finite cyclic group with generator $g$, and let  $x \in \mathbb{Z}$  is Alice’s private key. The element $g^x$ is public. To send a message $m \in  G,$ Bob picks a random   integer $y$ and sends the cipher text $c = (g^y,g^{xy}m)$ to Alice. To decrypt, Alice calculates $(g^y)^x=g^{xy}$ and inverts it to retrieve $m.$ There are a couple of cryptosystem of the ElGamal-type.  See, for example, \cite{KK}, \cite{FMR}. The versions proposed in \cite{Mah1} and \cite{Mah2} were analyzed in \cite{RO}. See also cryptanalysis in \cite{Romessays}.

 We are to show that the version of MOR in \cite{BMSS}  is not entirely accurate. It should be supplemented with an additional assumption.  The equivalence theorem there should  be clarified too. 
 
 We also show that the proposed ElGamal-type version of MOR over any finitely generated matrix group $G\leq $ GL($d, \mathbb{F}_q$)        is vulnerable with respect to the linear decomposition attack in any case when the automorphism $\varphi$ can be naturally extended to a linear transformation of the linear space Lin$_{\mathbb{F}_q}(G)$ generated by $G$ in M($d, \mathbb{F}$). For example, if  $\varphi$ is an inner automorphism.  
 In fact, there exists an efficient algorithm to compute  the original message by the its ciphertext. It can be done for every constructive field, i.e., a field for which all operations are efficient, and the Gauss elimination process is efficient too. 
 
\section{ Description of the ElGamal version of the MOR cryptosystem in \cite{BMSS}.} Let $G = <g_1,g_2, ..., g_n>$  be a (finite) public group and $\varphi $ a non-trivial public automorphism of $G$. 

Alice’s keys are as follows:

Private Key: $t, t\in \mathbb{N}.$  

Public Key: $(\{\varphi (g_i):  i=1, ..., n\}\ {\rm  and}\  \{\varphi^t(g_i): i=1, ..., n\}).$

We  suppose that Alice is the recipient of the messages and Bob is communicating with Alice. Let $m \in  G$ be a message.

Algorithm: 

Encryption.

To send the message (plaintext) $m$ Bob picks up a random integer $r$, then he computes $\{\varphi^r(g_i): i = 1, ..., n\}$ and $\varphi^{tr}(m)$. 

 The ciphertext is ($\{\varphi^r(g_i), i=1, ..., n\}, \varphi^{tr}(m)).$

Decryption.

Since Alice knows $t$, she  computes $\varphi^{tr}(g_i)$ from $\varphi^{r}(g_i)$ and then $\varphi^{-tr}(g_i)$\  ($i=1, ..., n$). Finally, the message $m$ can be  computed  by $\varphi^{-tr}(\varphi^{tr}(m))=m$.

\begin{remark}
\label{re:1}
There is one obstacle to the implementation of the decryption process.
To recover $m$, Alice should to compute  $\{\varphi^{-tr}(g_i): i = 1, ..., n\}$ by  $\{\varphi^{tr}(g_i): i =1, ..., n\}$,  or to compute it by $\varphi^r$. It can be done if she knows $\varphi^{-1}$, i.e., $\{\varphi^{-1}(g_i): i = 1, ..., n\}.$ 

In the general case, the calculation of the inverse automorphism is not  obvious efficient process. We have to assume that Alice can do it, for example, because she knows $s \in \mathbb{N}$ such that $\varphi^s=id.$ It happens, in particular,  if  she knows  the order $s_1$ of  $\varphi$ or  the order $s_2$    of  Aut($G$). Then $\varphi^{-1}=\varphi^{s-1}$ \ ($s = s_1$ or $s = s_2$).  Also Alice can know $\varphi^{-1}.$  

Alice can simultaneously build $\varphi$ and $\varphi^{-1}$ during the setting of parameters of the protocol. 
\end{remark}

This obstacle manifests itself more significantly in the proof of the following theorem. We give them in the original form.

\medskip
''{\bf Theorem} (\cite{BMSS}, Theorem 2.1). 
\label{th:1} The hardness to break the above MOR cryptosystem is equivalent to the Diffie-Hellman problem in the group  $<\varphi >$.

\medskip
{\it Proof}.
It is easy to see that if one can break the Diffie-Hellman problem, then one can compute $\varphi^{tr}$  from $\varphi^t$ in the public-key and $\varphi^r$ in the ciphertext. This breaks the system.

On the other hand, observe that the plaintext is $m=\varphi^{-tr}(\varphi^{tr}(m)).$ Assume that there is an oracle that can break the MOR cryptosystem, i.e., given $\varphi , \varphi^{t}$ and a ciphertext $(\varphi^r, f)$ will deliver $\varphi^{-tr}(f).$ Now we query the oracle $n$  times with the public-key and the ciphertexts ($\varphi^r(g_i), g_i$)      for $i = 1, ..., n$. From the output, one can easily find $\varphi^{-tr}(g_i)$ for $i=1, 2, ..., n.$  So we just witnessed that for $\varphi^r(g_i)$ and $\varphi^t(g_i)$ for $i=1, ..., n$, one can compute $\varphi^{-tr}(f)$  for every $f=f(g_1, ..., g_n)$ using the oracle. This solves the Diffie-Hellman problem.'' 

\begin{remark}
\label{re:2}
In the first part of the proof one computes $\varphi^{tr}$, but he needs in    $\varphi^{-tr}$ to compute $m$ in the protocol. It can be not so easy to do. There are some cryptographic schemes based on the complexity of the problem of finding the inverse to a given automorphism. 
\end{remark}

\section{Cryptanalysis}

We propose the following cryptanalysis that  works in the case of an arbitrary (constructive) field too. For simplicity we assume that the underlined field is finite. 

Suppose that the  ElGamal-type  system MOR  is considered over a finitely generated matrix group 
$G \leq $ GL($d, \mathbb{F}_q$).  Then $G \subseteq $ M($d, \mathbb{F}_q$). Let $G = <g_1, ..., g_n>.$ We suppose that $\varphi$ can be naturally extended to a linear transformation of $V=$Lin$_{\mathbb{F}_q}(G)$
that is a linear subspace generated by $G$ in  M($d,\mathbb{F}_q$). It happens for example, if $\varphi$ is an inner automorphism of $G.$ Note, that the case of inner automorphism $\varphi$ is considered in \cite{BMSS} as the most significant.

 To reveal $m$ using only open protocol data, we perform the following actions.
 
 Step 1. Let $V_i$ \  ($i \in \{1, ..., n\}$) be the subspace of $V$ generated by all elements of the form $\varphi^k(g_i)$ for $k \in \mathbb{Z}$. There is a basis of $V_i$ of the form $e_1(i)=\varphi^0(g_i) = g_i, e_2(i)=\varphi (g_i), ..., e_{l_i}(i)=\varphi^{l_i-1}(g_i).$ It can be efficiently constructed as follows. 
 
Initially, we include $e_1(i)=g_i$ in the constructing basis. Then we check whether $\varphi (g_i)$   belongs to the linear subspace generated by $e_1(i)$. 
If not, then we add $e_2(i)=\varphi (g_i)$  to the basis under construction.  
Let $e_1(i), ..., e_j(i)$ is a constructed part of the basis.  Then we check whether 
$\varphi^{j}(g_i) = \varphi (e_j(i))$   belongs to the linear subspace generated by $e_1(i), ..., e_j(i)$. If not, then we add $e_{j+1}(i)=\varphi^{j} (g_i)$  to the basis under construction, and continue. If so, we stop the process and claim that the basis is constructed and $l_i=j.$ Indeed, a linear presentation of $\varphi^{j}(g_i)$ via $e_1(i), ..., e_j(i)$ after applying $\varphi$ gives a linear presentation of $\varphi^{j+1}(g_i)$ via $e_2(i), ..., e_j(i), \varphi^j(g_i)$, and so via $e_1(i), ..., e_j(i)$. This argument works for every $j+v, v \geq 1$. Similarly we can obtain the linear decomposition of each $\varphi^{-v}(g_i), v \geq 1.$ 
 
 Step 2. For each $i = 1, ..., n,$ we have constructed a    basis $e_1(i), ..., e_{l_i}(i)$, where $e_{j+1}(i) = \varphi^{j}(g_i), j = 0, ..., l_i-1,$ of $V_i$. Each subspace $V_i$ is $\varphi$-invariant.  In general case $l_i\leq d^2$. 

  In \cite{BMSS},  the authors single out as the main the case of inner automorphism $\varphi .$ 
  
 They write:  
 
 ''The purpose of this section is to show that for a secure MOR cryptosystem over the classical Chevalley and twisted orthogonal groups, we have to look at automorphisms that act by conjugation like the inner automorphisms. There are other automorphisms that also act by conjugation, like the diagonal automorphism and the graph automorphism for odd-order orthogonal groups. Then we argue what is the hardness of our security assumptions.''

 Then they note that by Dieudonne Theorem, $\varphi  = \sigma \iota \eta \gamma \theta $, where  $\sigma$  is a central automorphism, $\iota$ is an inner automorphism, $\eta$ is a diagonal automorphism, $\gamma$  is a graph automorphism, and $\theta$ is a field automorphism. 
 
 Then they continue:
 
 ''The group of central automorphisms is too small and the field automorphisms reduce to a discrete logarithm in the field $F_q.$ So there is no benefit of using these in a MOR cryptosystem. Also there are not many graph automorphisms in classical Chevalley and twisted orthogonal groups other than special linear groups and oddorder orthogonal groups. In the odd-order orthogonal groups, these automorphisms act by conjugation. Recall here that our automorphisms are presented as action on generators. It is clear (\cite{Mah2}, Section 7) that if we can recover the conjugating matrix from the action on generators, the security is a discrete logarithm problem in $\mathbb{F}_{q^{d}}$, or else the security is a discrete logarithm problem in $F_{q^{d^2}}$.'' 
 
 In our cryptanalysis we suppose that $\varphi$ can be naturally extended to an automorphism of the linear space $V$.  This happens if $\varphi$ is an inner or field automorphism, or induced by an inner automorphism of GL($d, \mathbb{F}_q$).

 We come back to the introduced above subspaces $V_i, i = 1, ..., n.$ For a fixed $V_i$, denote by $\varphi_i$ the linear map of $V_i$ induced by $\varphi$. The matrix $A(\varphi_i)$ of $\varphi_i$ in the basis $E_i=
 \{e_1(i), ..., e_{l_i}(i)\}$ has the form 
 
 $$
 A(\varphi_i) = \left(\begin{array}{cccccc}
 0&1&0&...&...&0\\
 0&0&1&0&...&0\\
 ...&...&...&...&...&...\\
 0&...&...&...&0&1\\
 \alpha_1& \alpha_2&...&...&...&\alpha_{l_i}\\
 \end{array}\right),
 $$
 \noindent
 where $\varphi (e_{l_i}(i))= \sum_{k=1}^{l_i}\alpha_ke_{k}(i), \alpha_k\in \mathbb{F}_q.$ 
 
 By the way we can efficiently compute for each $i$ the value $\varphi^{-1}(g_i)$ corresponding to the first row of $A(\varphi_i)^{-1}.$ So we can compute $\varphi^{-1}.$ 
 
 Now we know matrices $A(\varphi_i)^{\pm 1}, A(\varphi_i)^{\pm r}, A(\varphi_i)^{\pm t}, i=1, ..., n,$ and we need to compute $r$ or $t$. Then we can compute $\varphi_i^{- rt}$ and recover $m.$ We can provide sufficient computation using only one or several matrices above.

In \cite{MV}, it was shown how the discrete logarithm problem in some special class of matrices can be reduced to the discrete logarithm problem in some extensions of the underlying field. In \cite{MW}, these results were extended to show how the discrete logarithm problem in every group GL($d, \mathbb{F}_q$) can be reduced in probabilistic polynomial time to the similar problem in small extensions of $\mathbb{F}_q$. The case of a finitely generated nilpotent group is considered in \cite{Romnil}.

  We see that matrix groups over finite fields offer no significant advantage for the implementation of cryptographic protocols whose security is based on the difficulty of computing discrete logarithms.

  The described cryptanalysis has many analogues, presented in \cite{Romalg}-\cite{MR}. In \cite{Romgen}, a general scheme based on multiplications is presented. It corresponds to a number of cryptographic systems known in the literature, which are also vulnerable to attacks by the linear decomposition method. The nonlinear decomposition method was invented in \cite{Romnon}. The nonlinear method can be applied when the group chosen as the platform for a cryptographic scheme is not linear or the least degree of their representability by matrices is too big for efficient computations. See details in \cite{Romessays}. 
  
  A protection against linear algebra attacks was recently  invented in \cite{R5}. It is described in the case of the Anshel et al. cryptographic scheme \cite{AAG} but can be applied to the Diffie-Hellman-type and some other schemes too. See abstract \cite{RompdAAG} and paper \cite{RomVestnikAAG}.

\end{document}